\documentstyle[preprint,eqsecnum,epsf,aps,floats,amssymb,tighten]{revtex}  
\begin{document}
\def\theequation{\arabic{equation}}%
\newcommand{\rstev}{\mbox{$\rs = \T{1.8}$}}
\newcommand{\XX}{\mbox{$\, \times \,$}}
\newcommand{\AP}{\mbox{${\rm \bar{p}}$}}
\newcommand{\SU}{\mbox{$S$}}
\newcommand{\SPt}{\mbox{$<\! |S|^2 \!>$}}
\newcommand{\ET}{\mbox{$E_{T}$}}
\newcommand{\PT}{\mbox{$p_{t}$}}
\newcommand{\DP}{\mbox{$\Delta\phi$}}
\newcommand{\DR}{\mbox{$\Delta R$}}
\newcommand{\DE}{\mbox{$\Delta\eta$}}
\newcommand{\DEP}{\mbox{$\Delta\eta_{c}$}}
\newcommand{\DEC}{\mbox{$\Delta\eta_{c}$}}
\newcommand{\SP}{\mbox{$S(\DEP)$}}
\newcommand{\PH}{\mbox{$\phi$}}
\newcommand{\EA}{\mbox{$\eta$}}
\newcommand{\EAJ}{\mbox{\EA(jet)}}
\newcommand{\AEA}{\mbox{$|\eta|$}}
\newcommand{\Ge}[1]{\mbox{#1 GeV}}
\newcommand{\T}[1]{\mbox{#1 TeV}}
\newcommand{\D}[1]{\mbox{$#1^{\circ}$}}
\newcommand{\x}{\cdot}
\newcommand{\ra}{\rightarrow}
\newcommand{\mb}{\mbox{mb}}
\newcommand{\nb}{\mbox{nb}}
\newcommand{\ipb}{\mbox{${\rm pb}^{-1}$}}
\newcommand{\inb}{\mbox{${\rm nb}^{-1}$}}
\newcommand{\rs}{\mbox{$\sqrt{s}$}}
\newcommand{\fdel}{\mbox{$f(\DEP)$}}
\newcommand{\fdele}{\mbox{$f(\DEP)^{exp}$}}
\newcommand{\fgap}{\mbox{$f(\DEP\! > \!3)$}}
\newcommand{\fgape}{\mbox{$f(\DEP\! > \!3)^{exp}$}}
\newcommand{\fpyt}{\mbox{$f(\DEP\!>\!2)$}}
\newcommand{\delth}{\mbox{$\DEP\! > \!3$}}
\newcommand{\uplim}{\mbox{$1.1\!\times\!10^{-2}$}}
\newcommand{\sigew}{\mbox{$\sigma_{\rm EW}$}}
\newcommand{\sigsi}{\mbox{$\sigma_{\rm singlet}$}}
\newcommand{\sigr}{\mbox{$\sigsi/\sigma$}}
\newcommand{\sigrew}{\mbox{$\sigew/\sigma$}}
\newcommand{\ncal}{\mbox{$n_{\rm cal}$}}
\newcommand{\ntrk}{\mbox{$n_{\rm trk}$}}

\def\simge
{\mathrel{\rlap{\raise 0.53ex \hbox{$>$}}{\lower 0.53ex \hbox{$\sim$}}}}

\def\simle
{\mathrel{\rlap{\raise 0.4ex \hbox{$<$}}{\lower 0.72ex \hbox{$\sim$}}}}

%
\def\sigtot{$\sigma_{\rm tot}$}         
\def\sigtop{$\sigma_{t \overline{t}}$}  
\def\pbarp{$\overline{p}p $}            
\def\ppbar{$p\overline{p} $}            
\def\qqbar{$q\overline{q}$}             
\def\ttbar{$t\overline{t}$}             
\def\bbbar{$b\overline{b}$}             
\def\D0{D\O}                            
\def\ipb{pb$^{-1}$}                     
\def\pt{p_T}                            
\def\ptg{p_T^\gamma}                    
\def\et{E_T}                            
\def\etg{E_T^\gamma}                    
\def\htran{$H_T$}                       
\def\gevcc{{\rm GeV}/c$^2$}                   
\def\gevc{{\rm GeV}/c}                  
\def\gev{\rm GeV}                       
\def\tev{\rm TeV}                       
\def\njet{$N_{\rm jet}$}                
\def\aplan{$\cal{A}$}                   
\def\lum{$\cal{L}$}                     
\def\iso{$\cal{I}$}                     
\def\remu{${\cal{R}}_{e\mu}$}           
\def\rmu{$\Delta\cal{R}_{\mu}$}         
\def\pbar{$\overline{p}$}               
\def\tbar{$\overline{t}$}               
\def\bbar{$\overline{b}$}               
\def\lumint{$\int {\cal{L}} dt$}        
\def\lumunits{cm$^{-2}$s$^{-1}$}        
\def\etal{{\sl et al.}}                 
\def\vs{{\sl vs.}}                      
\def\sinthw{sin$^2 \theta_W$}           
\def\mt{$m_t$}                          
\def\mb{$m_b$}                          
\def\mw{$M_W$}                          
\def\mz{$M_Z$}                          
\def\pizero{$\pi^0$}                    
\def\jpsi{$J/\psi$}                     
\def\wino{$\widetilde W$}               
\def\zino{$\widetilde Z$}               
\def\squark{$\widetilde q$}             
\def\gluino{$\widetilde g$}             
\def\alphas{$\alpha_{\scriptscriptstyle S}$}                
\def\alphaem{$\alpha_{\scriptscriptstyle{\rm EM}}$}         
\def\epm{$e^+e^-$}                      
\def\deg{$^\circ$}                      
\def\met{\mbox{${\hbox{$E$\kern-0.6em\lower-.1ex\hbox{/}}}_T$ }} 

\def\disc{\mbox{$\log_{10}\left[{1+\log_{10}{\{1+E_1({\rm GeV})\}}}\right]$}}

\newcommand{\NC}{{\em Nuovo Cimento\/} }
\newcommand{\NIM}{{\em Nucl. Instr. Meth.} }
\newcommand{\NP}{{\em Nucl. Phys.} }
\newcommand{\PL}{{\em Phys. Lett.} }
\newcommand{\PR}{{\em Phys. Rev.} }
\newcommand{\PRL}{{\em Phys. Rev. Lett.} }
\newcommand{\RMP}{{\em Rev. Mod. Phys.} }
\newcommand{\ZP}{{\em Zeit. Phys.} }
\def\err#1#2#3 {{\it Erratum} {\bf#1},{\ #2} (19#3)}
\def\ib#1#2#3 {{\it ibid.} {\bf#1},{\ #2} (19#3)}
\def\nc#1#2#3 {Nuovo Cim. {\bf#1} ,#2(19#3)}
\def\nim#1#2#3 {Nucl. Instr. Meth. {\bf#1},{\ #2} (19#3)}
\def\np#1#2#3 {Nucl. Phys. {\bf#1},{\ #2} (19#3)}
\def\pl#1#2#3 {Phys. Lett. {\bf#1},{\ #2} (19#3)}
\def\prev#1#2#3 {Phys. Rev. {\bf#1},{\ #2} (19#3)}
\def\prl#1#2#3 {Phys. Rev. Lett. {\bf#1},{\ #2} (19#3)}
\def\rmp#1#2#3 {Rev. Mod. Phys. {\bf#1},{\ #2} (19#3)}
\def\zp#1#2#3 {Zeit. Phys. {\bf#1},{\ #2} (19#3)}

\hyphenation{PYTHIA}
\hyphenation{ident-if-ic-ation}
\hyphenation{mes-ons}
\hyphenation{reg-ions}
\hyphenation{weights}
\hyphenation{scale}
\hyphenation{pre-dicts}
\hyphenation{extends}
\lefthyphenmin=2 
\righthyphenmin=3

\parskip \baselineskip


\preprint{Fermilab-Pub-99/354-E}

\title{The Isolated Photon Cross Section in 
$p \overline p$ Collisions at \rstev \footnote{submitted to Physical
Review Letters}}

%
\author{                                                                      
B.~Abbott,$^{47}$                                                             
M.~Abolins,$^{44}$                                                            
V.~Abramov,$^{19}$                                                            
B.S.~Acharya,$^{13}$                                                          
D.L.~Adams,$^{54}$                                                            
M.~Adams,$^{30}$                                                              
S.~Ahn,$^{29}$                                                                
V.~Akimov,$^{17}$                                                             
G.A.~Alves,$^{2}$                                                             
N.~Amos,$^{43}$
% ...
\\                                                                            
\vskip 0.30cm                                                                 
\centerline{(D\O\ Collaboration)}                                             
\vskip 0.30cm                                                                 
}                                                                             
\address{                                                                     
\centerline{$^{1}$Universidad de Buenos Aires, Buenos Aires, Argentina}       
\centerline{$^{2}$LAFEX, Centro Brasileiro de Pesquisas F{\'\i}sicas,         
                  Rio de Janeiro, Brazil}                                     
\centerline{$^{3}$Universidade do Estado do Rio de Janeiro,                   
                  Rio de Janeiro, Brazil}                                     
\centerline{$^{4}$Institute of High Energy Physics, Beijing,                  
                  People's Republic of China}                                 
\centerline{$^{5}$Universidad de los Andes, Bogot\'{a}, Colombia}             
\centerline{$^{6}$Universidad San Francisco de Quito, Quito, Ecuador}         
\centerline{$^{7}$Institut des Sciences Nucl\'eaires, IN2P3-CNRS,             
                  Universite de Grenoble 1, Grenoble, France}                 
\centerline{$^{8}$Centre de Physique des Particules de Marseille,             
                  IN2P3-CNRS, Marseille, France}                              
\centerline{$^{9}$LPNHE, Universit\'es Paris VI and VII, IN2P3-CNRS,          
                  Paris, France}                                              
\centerline{$^{10}$DAPNIA/Service de Physique des Particules, CEA, Saclay,    
                  France}                                                     
\centerline{$^{11}$Panjab University, Chandigarh, India}                      
\centerline{$^{12}$Delhi University, Delhi, India}                            
\centerline{$^{13}$Tata Institute of Fundamental Research, Mumbai, India}     
\centerline{$^{14}$Seoul National University, Seoul, Korea}                   
\centerline{$^{15}$CINVESTAV, Mexico City, Mexico}                            
\centerline{$^{16}$Institute of Nuclear Physics, Krak\'ow, Poland}            
\centerline{$^{17}$Institute for Theoretical and Experimental Physics,        
                   Moscow, Russia}                                            
\centerline{$^{18}$Moscow State University, Moscow, Russia}                   
\centerline{$^{19}$Institute for High Energy Physics, Protvino, Russia}       
\centerline{$^{20}$Lancaster University, Lancaster, United Kingdom}           
\centerline{$^{21}$University of Arizona, Tucson, Arizona 85721}              
\centerline{$^{22}$Lawrence Berkeley National Laboratory and University of    
                   California, Berkeley, California 94720}                    
\centerline{$^{23}$University of California, Davis, California 95616}         
\centerline{$^{24}$California State University, Fresno, California 93740}     
\centerline{$^{25}$University of California, Irvine, California 92697}        
\centerline{$^{26}$University of California, Riverside, California 92521}     
\centerline{$^{27}$Florida State University, Tallahassee, Florida 32306}      
\centerline{$^{28}$University of Hawaii, Honolulu, Hawaii 96822}              
\centerline{$^{29}$Fermi National Accelerator Laboratory, Batavia,            
                   Illinois 60510}                                            
\centerline{$^{30}$University of Illinois at Chicago, Chicago,                
                   Illinois 60607}                                            
\centerline{$^{31}$Northern Illinois University, DeKalb, Illinois 60115}      
\centerline{$^{32}$Northwestern University, Evanston, Illinois 60208}         
\centerline{$^{33}$Indiana University, Bloomington, Indiana 47405}            
\centerline{$^{34}$University of Notre Dame, Notre Dame, Indiana 46556}       
\centerline{$^{35}$Purdue University, West Lafayette, Indiana 47907}          
\centerline{$^{36}$Iowa State University, Ames, Iowa 50011}                   
\centerline{$^{37}$University of Kansas, Lawrence, Kansas 66045}              
\centerline{$^{38}$Kansas State University, Manhattan, Kansas 66506}          
\centerline{$^{39}$Louisiana Tech University, Ruston, Louisiana 71272}        
\centerline{$^{40}$University of Maryland, College Park, Maryland 20742}      
\centerline{$^{41}$Boston University, Boston, Massachusetts 02215}            
\centerline{$^{42}$Northeastern University, Boston, Massachusetts 02115}      
\centerline{$^{43}$University of Michigan, Ann Arbor, Michigan 48109}         
\centerline{$^{44}$Michigan State University, East Lansing, Michigan 48824}   
\centerline{$^{45}$University of Nebraska, Lincoln, Nebraska 68588}           
\centerline{$^{46}$Columbia University, New York, New York 10027}             
\centerline{$^{47}$New York University, New York, New York 10003}             
\centerline{$^{48}$University of Rochester, Rochester, New York 14627}        
\centerline{$^{49}$State University of New York, Stony Brook,                 
                   New York 11794}                                            
\centerline{$^{50}$Brookhaven National Laboratory, Upton, New York 11973}     
\centerline{$^{51}$Langston University, Langston, Oklahoma 73050}             
\centerline{$^{52}$University of Oklahoma, Norman, Oklahoma 73019}            
\centerline{$^{53}$Brown University, Providence, Rhode Island 02912}          
\centerline{$^{54}$University of Texas, Arlington, Texas 76019}               
\centerline{$^{55}$Texas A\&M University, College Station, Texas 77843}       
\centerline{$^{56}$Rice University, Houston, Texas 77005}                     
\centerline{$^{57}$University of Washington, Seattle, Washington 98195}       
}                                                                             

\date{\today}

\maketitle

\begin{abstract}
{\baselineskip 24pt
We report a new measurement of the cross section for the production of 
isolated photons, with transverse energies ($\etg$) above 10~GeV 
and pseudorapidities $|\eta| < 2.5$,
in $p\bar{p}$ collisions at $\sqrt{s} = 1.8\,$TeV.
The results are based on a data sample of 107.6~pb$^{-1}$ recorded during
1992--1995 with the \D0\ detector at the Fermilab Tevatron collider.
The background, predominantly from jets which fragment to neutral
mesons, was estimated using the longitudinal shower shape of photon 
candidates in the calorimeter.   
The measured cross section is in good agreement with
the next-to-leading order (NLO) 
QCD calculation for $\etg\ \simge 36\,{\rm GeV}$.}
\end{abstract}
\vspace{-1.cm}

\vspace{-1.cm}


Direct (or prompt) photons, by which we mean those produced 
in a hard parton-parton interaction, provide a probe of
the hard scattering process which 
minimizes confusion from parton fragmentation or from experimental issues
related to jet identification and energy measurement~\cite{farrar}.
In high energy $p \overline p$ collisions
the dominant mode for production of
photons with moderate transverse energy $\etg$ is
through the strong Compton process $qg \to q\gamma$.  
The direct photon cross section is thus
sensitive to the gluon distribution in the proton. 
Direct-photon measurements 
allow tests of NLO and resummed QCD calculations, phenomenological
models of gluon radiation, and studies of photon isolation
and the fragmentation process. 

Data from previous collider measurements~\cite{uatwo,cdf,prl1a}
have indicated an excess of photons at low
$\etg\ (\simle 25\,{\rm GeV})$ compared with predictions
of NLO QCD.
This excess may originate in 
additional gluon radiation beyond that included in the QCD 
calculation~\cite{kt}, or reflect inadequacies in the parton
distributions and fragmentation contributions~\cite{vvv}.

In this Letter, we present a new measurement of 
the cross section for production of isolated photons
with $\etg \geq 10$~GeV and pseudorapidity $|\eta|< 2.5$
in $p\overline p$ collisions at $\sqrt{s} = 1.8\,$TeV, which supersedes
our previous publication~\cite{prl1a}. 
(Pseudorapidity is defined as
$\eta = - {\rm ln\,tan}{\theta \over 2}$ where
$\theta$ is the polar angle with respect to the proton beam.)
The higher statistical
precision afforded by the increased luminosity
($12.9\pm 0.7$~pb$^{-1}$ recorded during 1992--1993 and 
$94.7\pm 5.1$~pb$^{-1}$ recorded during 1994--1995) motivated
a refined estimation of the backgrounds. In particular, 
fully-simulated jet events were used in place of single neutral 
mesons to model background.  

Photon candidates were identified in the \D0\ detector~\cite{dzero} as
isolated clusters of energy depositions in the uranium and liquid-argon 
sampling calorimeter.  The calorimeter covered $|\eta| \simle 4$ and had 
electromagnetic (EM) energy resolution
$\sigma_E/{E} \approx 15\%/\sqrt{E (\rm GeV)} \oplus 0.3\%$.
The EM section of the calorimeter was segmented longitudinally into 
four layers (EM1--EM4) of 2, 2, 7, and 10 radiation lengths respectively, 
and transversely into cells in pseudorapidity and azimuthal angle 
$\Delta \eta \times \Delta \phi = 0.1 \times 0.1$ ($0.05 \times 0.05$ at shower
maximum in EM3). Drift chambers in front of the calorimeter were used 
to distinguish photons from electrons, or from photon conversions, 
by ionization measurement. 

A three-level trigger was employed during data taking.
The first level used scintillation counters near the 
beam pipe to detect an inelastic interaction;
the second level required that the EM energy in calorimeter towers of size
$\Delta \eta \times \Delta \phi =0.2 \times 0.2$ be above
a programmable threshold. The third level was a software 
trigger in which clusters of calorimeter cells were required
to pass minimal criteria on shower shape.

Offline, candidate clusters were accepted within
the regions $| \eta | < 0.9 $ (central) and
$1.6 < |\eta|< 2.5$ (forward) to avoid inter-calorimeter boundaries; in the 
central region, clusters were required to be more than 1.6~cm 
from azimuthal boundaries of modules.
The event vertex was required to be within 50~cm of the nominal 
center of the detector along the beam.
Each candidate was required to have a shape consistent with that
of a single EM shower, to deposit more than 96\% of the energy detected
in the calorimeter in the EM section, and to be isolated as defined
by the following requirements on the transverse energy 
observed in the annular region between
${\cal R}=\sqrt{\Delta\eta^2 + \Delta\phi^2}=0.2$ and ${\cal R}=0.4$ 
around the cluster:
$E_T^{{\cal R}\leq 0.4} - E_T^{{\cal R}\leq 0.2} < 2\,$GeV. 
The combined
efficiency of these selections was estimated as a function of $\etg$ 
using a detailed Monte Carlo simulation of the detector~\cite{d0geant} 
and verified with electrons from $Z \to ee$ events, and found to be 
$0.65 \pm 0.01~(0.83\pm0.01)$
at $\etg=40\,{\rm GeV}$ for central (forward) photons.  
An uncertainty of 2.5\%  was added in quadrature to this
to allow for a possible dependence on instantaneous luminosity.
Photon candidates were rejected if there were tracks within a road
$\Delta\theta \times\Delta\phi
\approx 0.2\times 0.2$ radians between the calorimeter cluster and
the primary vertex. The mean efficiency of this requirement was 
measured to be $0.83\pm 0.01~(0.54\pm 0.03)$ in the
central (forward) region. The inefficiency stemmed mainly from
photon conversions and overlaps of photons with charged tracks
(either from the underlying event or from other $p \overline p$ interactions).

Background to the direct-photon signal comes primarily
from two-photon decays of $\pi^0$ and $\eta$ mesons produced in jets. 
While the bulk of this background is rejected by the 
selection criteria (especially the isolation requirement),
substantial contamination remains,
predominantly from fluctuations in jet fragmentation, which can 
produce neutral mesons that carry most of the jet energy.  
For a $\pi^0$ meson with $\etg \simge 10\,$GeV, 
the showers from its two-photon decay
coalesce and mimic a single photon in the calorimeter.  

\begin{figure}[t]
\begin{center}
\epsfxsize=4in
\leavevmode\epsffile{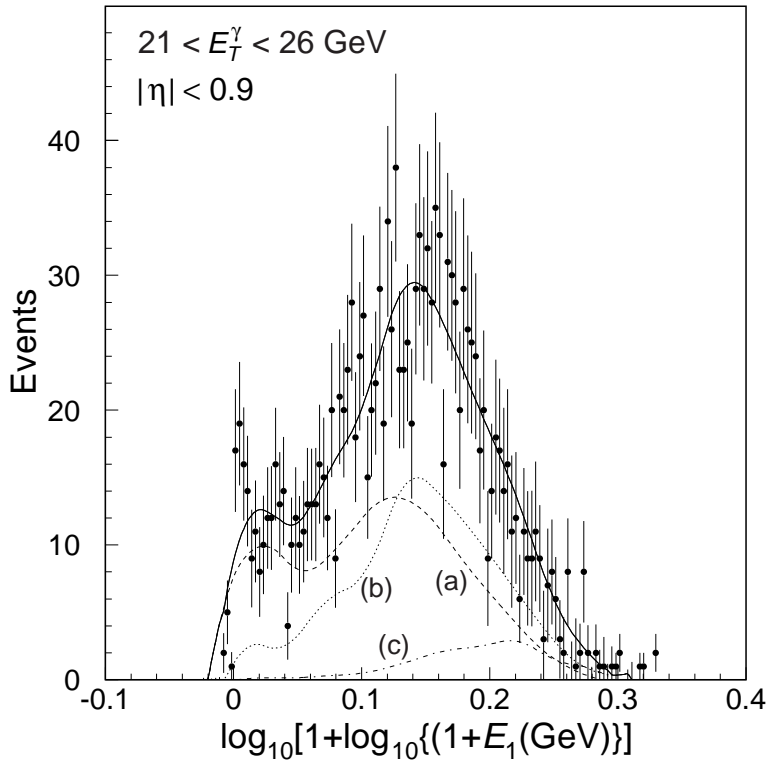}
\vspace{12pt}
\caption{Distribution of the discriminant variable for 
$21< \etg< 26\,{\rm GeV}$ central photon candidates (points with error bars),
and the fitted distribution (solid curve) composed of Monte Carlo
photons (curve labelled $(a)$) and jets with and without charged particles 
(curves labelled $(c)$ and $(b)$, respectively).  
The Monte Carlo curves shown here 
were smoothed for clarity (this was not done in the fitting itself). 
Results of these fits provide the 
purity  ${\cal P}$ of the signal: for this bin, ${\cal P}= 0.58\pm0.07$.
\label{emf}}
\end{center}
\end{figure}

The fraction of the remaining candidates 
that are genuine direct photons (the purity ${\cal P}$) was 
determined using the energy $E_1$ deposited in the first 
layer (EM1) of the calorimeter. 
The decays of neutral mesons primarily responsible for background 
produce two nearby photons, and the probability that at least one of 
them undergoes a conversion to an $e^+e^-$ pair either in the cryostat 
of the calorimeter or the first absorber plate
is roughly twice that for a single photon. Such showers due to meson 
decays therefore start earlier in the calorimeter than 
showers due to single photons, and yield larger $E_1$ depositions 
for any initial energy.  A typical distribution in our discriminant, 
{\disc}, is shown in Fig.~\ref{emf}.  This variable emphasized 
differences between direct photons and background, and was insensitive to 
noise and event pileup.
A small correction, based on electrons from $W$ decays,
was made to bring the $E_1$ distribution for the 1992--1993 data 
into agreement with the 1994--1995 data.   
The distribution in the discriminant was then fitted to the sum of a 
photon signal and jet background, both of which were obtained
from Monte Carlo simulation.  
Two components of the jet background
were included separately: those with and those without charged tracks 
inside the inner isolation cone (${\cal R}=0.2$ from the photon candidate).
This was done to minimize constraints in the fit from the 
(relatively poorly determined) 
tracking efficiency and from the model used for jet fragmentation. 

\begin{figure}[t]
\begin{center}
\epsfxsize=4in
\leavevmode\epsffile{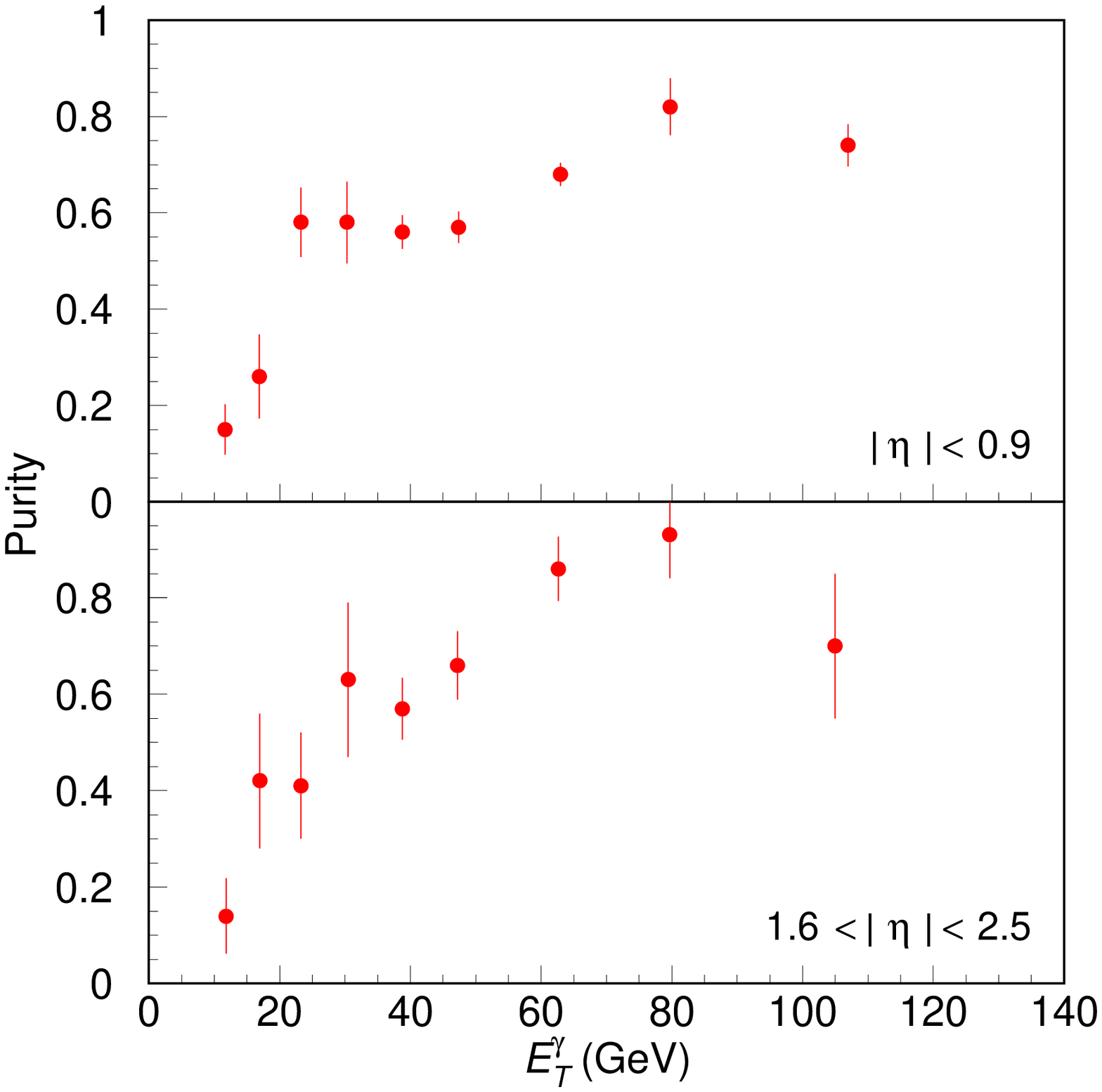}
\vspace{12pt}
\caption{
The fraction of photon candidates that are direct photons
as a function of $E_{T}^{\gamma}$, for central and forward photons.
\label{pur}}
\end{center}
\end{figure}

Direct photon and QCD jet events were 
generated using {\sc pythia}~\cite{pythia} and 
then passed through the {\sc geant} detector-simulation package, and
overlaid with data acquired using a random trigger to  
model noise, pileup, underlying event, and multiple $p \overline p$
interactions~\cite{d0geant}.  
The simulated $E_1$ was corrected for imperfect modeling
of the material in the detector.
We assumed that the Monte Carlo energy could be parametrized as
$E_1^{\rm MC} = \alpha + \beta E_1$, with the parameters $\alpha$ and $\beta$
determined from data:  $\beta$ from the $W \to e \nu$ sample and
$\alpha$ from the photon data. The fits to extract the purity ${\cal P}$
were performed for different values of $\alpha$, and the total $\chi^2$ 
was minimized for all $\etg$.
\begin{figure}[tb]
\begin{center}
\epsfxsize=4in
\leavevmode\epsffile{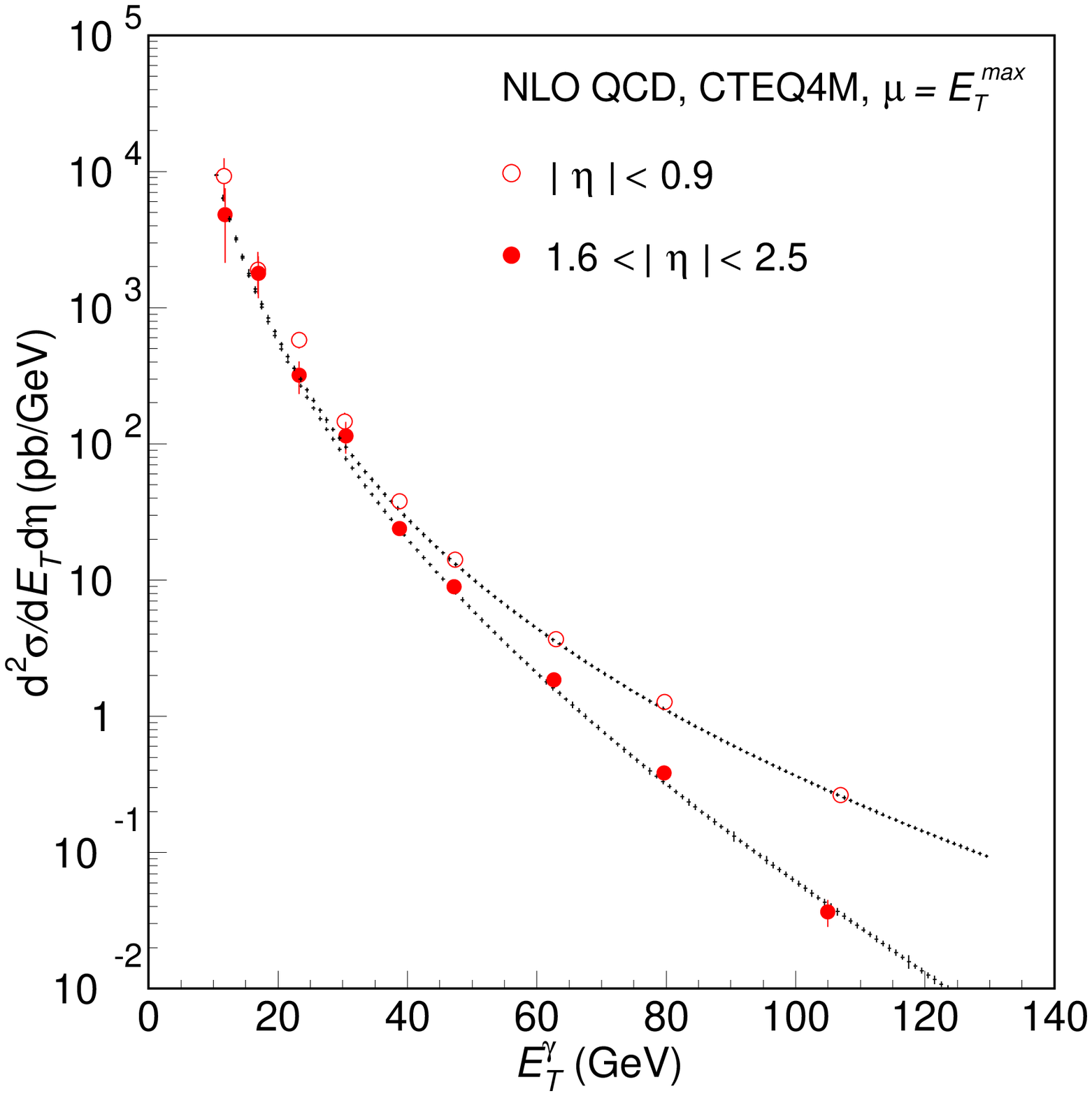}
\vspace{12pt}
\caption{
The cross section 
$d^2\sigma/dE_T^\gamma\,d\eta$ for isolated photons
as a function of transverse energy $\etg$, for central
and forward regions. The curves show the NLO QCD calculated cross
sections.
\label{log}}
\end{center}
\end{figure}

\begin{figure}[htb]
\begin{center}
\epsfxsize=4in
\leavevmode\epsffile{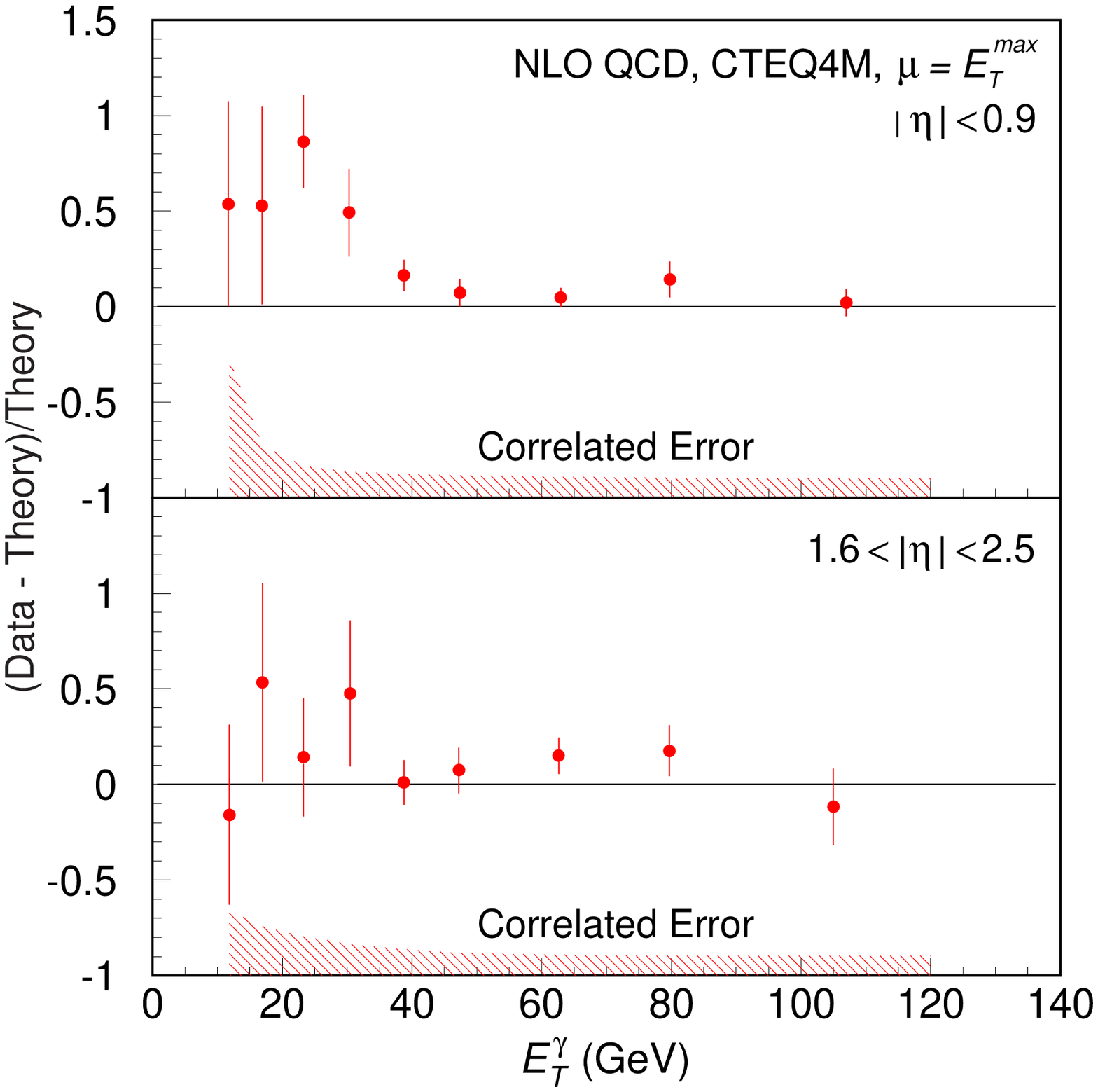}
\vspace{12pt}
\caption{Difference between the measured differential cross section
for isolated photon production
and the prediction from NLO QCD, using {\mbox CTEQ4M} parton distributions. 
\label{lin}}
\end{center}
\end{figure}

To reduce computation time, the jet background events were preselected 
just after their generation to have highly electromagnetic jets.
The background subtraction technique used in this analysis employs
fully-simulated jet events, whereas the previous analysis modeled 
the background with isolated neutral mesons.  With our increased statistics, 
it was found that individual isolated mesons could not adequately
model the background.  Indeed, our simulation shows that less
than half of the background can be attributed to the presence of
single neutral mesons within the inner isolation cones (of ${\cal R}=0.2$). 
The new approach provided a much better description of the shower shape 
and isolation energy, and resulted in an increased estimate of the signal 
purity.  

Fitting was done separately for samples at central and forward regions,
for each $\etg$ bin, using the package {\sc hmcmll}~\cite{hmcmll}, 
with the constraint that the fractions of signal and background were 
between 0.0 and 1.0. The resulting purity ${\cal P}$ and its uncertainty
is shown in Fig.~\ref{pur} as a function of $\etg$.  As well as the
fitting error, a systematic error was assigned to the use of
{\sc pythia} to model jets. This uncertainty
was estimated by varying the multiplicity of neutral mesons in the
core of the jet by $\pm 10$\%~\cite{fragerror}.

The differential cross section $d^2\sigma/d\etg\,d\eta$, determined
after correction for purity and efficiency (but not corrected
for energy resolution) is shown as a 
function of $\etg$ in Fig.~\ref{log} and in Table~\ref{tableone}.
The purity corrections were applied point by point,
using the same binning for the cross section as for the determination of
purity.  The correlated errors consist of the quadrature sum
of the uncertainties on luminosity, vertex requirements, 
and energy scale in the Monte Carlo  
(which are energy independent) and the model for fragmentation 
(large uncertainty at low $\etg$ because of the low purity in this region).   
The uncorrelated errors include the statistical uncertainty, the fitting
error, and the statistical
uncertainties on the determination of acceptance,
trigger efficiency, and the efficiency of the selection criteria.
 
These new measurements are $\approx 20-30$\% higher than
our previously published results. The change is well understood, and is
due to the improvements in the Monte Carlo model used to estimate the purity, 
and in calculations of the acceptance and luminosity\cite{bantly}.

We compare the measured cross section with NLO QCD calculations using
the program of Baer, Ohnemus, and Owens~\cite{owens}.  
This calculation includes $\gamma+{\rm jet}$,
$\gamma+{\rm two}$ jets, and two jets with bremsstrahlung in the final state.  
In the latter case, a jet collinear with the photon  was created with the 
remaining fraction of the energy of the relevant final-state parton, so that 
the isolation cut could be modeled.
For all sources of signal, the final-state parton energies were smeared using 
the measured EM and jet resolutions.  
The isolation criterion was imposed by rejecting events 
with a jet of $E_T > 2\,{\rm GeV}$ within ${\cal R}\le0.4$ of the photon.  
(Smearing photon and jet energies changed the QCD prediction by less than 4\%.)
CTEQ4M parton distributions~\cite{CTEQ} were used in the NLO calculations, 
with renormalization and factorization scales $\mu_R=\mu_F=E_T^{\rm max}$, 
where $E_T^{\rm max}$ is the larger of the transverse
energies of the photon or the leading jet. If, instead, the scales
$\mu_R=\mu_F=2E_T^{\rm max}$ or $E_T^{\rm max}/2$ were employed,
the predicted cross sections changed by $\simle 6$\%. 

Figure~\ref{lin} shows 
the difference between experimental and theoretical differential
cross sections ($d^2\sigma/d\etg\,d\eta$), divided by the theoretical values.
In both central and forward regions, the NLO QCD predictions agree with 
the data for transverse energies $\etg \simge 36\,{\rm GeV}$.
At lower transverse energies, particularly for $|\eta| < 0.9$,
our measured cross section exceeds the expectation from NLO QCD,
a trend consistent with previous observations at
collider~\cite{uatwo,cdf,prl1a} and fixed target~\cite{e706} energies. 
Using contributions from both correlated and uncorrelated errors, 
the $\chi^2$ value for the data compared with NLO QCD is 8.9 in the 
central region and 1.9 in the forward region, for 
$\etg \leq 36\,{\rm GeV}$ in each case (the first 4 data points).  

These data complement and extend 
previous measurements, and 
provide additional input
for extraction of parton distributions through global fits to
all data. The difference between the data and NLO QCD for $\etg \simle 36$
GeV suggests that a more complete theoretical understanding of processes that
contribute to the low-$\etg$ behavior of the photon cross section is needed.

%

We thank J.~F.~Owens for his assistance with the theoretical calculations.
%
We thank the Fermilab and collaborating institution staffs for 
contributions to this work, and acknowledge support from the 
Department of Energy and National Science Foundation (USA),  
Commissariat  \` a L'Energie Atomique (France), 
Ministry for Science and Technology and Ministry for Atomic 
   Energy (Russia),
CAPES and CNPq (Brazil),
Departments of Atomic Energy and Science and Education (India),
Colciencias (Colombia),
CONACyT (Mexico),
Ministry of Education and KOSEF (Korea),
CONICET and UBACyT (Argentina),
A.P. Sloan Foundation,
and the Humboldt Foundation.
%

\begin{table}[p]
\begin{center}
\caption{The predicted and measured cross sections in bins of $\etg$.
$\langle\etg\rangle$ is the average photon transverse energy in each bin.
The columns labelled $\delta\sigma_U$ and $\delta\sigma_C$ show 
the magnitude of the uncorrelated and correlated uncertainties,
respectively. (The statistical error is contained in $\delta\sigma_U$.)
\label{tableone}}
\begin{tabular}{cccccc}
$\etg$ bin	& $\langle\etg\rangle$	
&\multicolumn{2}{c}{$d^2 \sigma/dE_T^\gamma d\eta$~(pb/GeV)}
                        & $\delta\sigma_U$ &$\delta\sigma_C$\\
(GeV)		& (GeV)	&NLO QCD&measured& (\%) & (\%)\\
\hline\hline
\multicolumn{6}{c}{$|\eta| < 0.9$} \\
\hline
  10.0 --  14.0 &   11.7 &  6030 & 9270 &     35 &     74  \\
  14.0 --  21.0 &   16.9 &  1250 & 1910 &     34 &     27  \\
  21.0 --  26.0 &   23.3 &  310  & 579 &     13 &     17  \\
  26.0 --  36.0 &   30.3 &  97.9 & 146 &     15 &     14  \\
  36.0 --  42.0 &   38.8 &  32.5 & 37.8 &      7.1 &     13  \\
  42.0 --  54.0 &   47.4 &  13.1 & 14.1 &      6.7 &     12 \\
  54.0 --  75.0 &   63.0 &  3.52 & 3.69 &      4.8 &     11  \\
  75.0 --  85.0 &   79.8 &  1.12 & 1.28 &      8.3 &     11  \\
  85.0 -- 140.0 &  106.8 &  0.258& 0.264 &      7.1 &     10  \\
\hline
\multicolumn{6}{c}{$1.6 < |\eta| < 2.5$}  \\
\hline
 10.0 --  14.0 &   11.8 & 5760  &4850 &     56 &     34  \\
 14.0 --  21.0 &   17.0 & 1160  &1780 &     34 &     26  \\
 21.0 --  26.0 &   23.3 & 279   &318 &     27 &     20  \\
 26.0 --  36.0 &   30.5 & 77.9  &115 &     26 &     17  \\
 36.0 --  42.0 &   38.8 & 23.6  &23.8 &     12 &     14  \\
 42.0 --  54.0 &   47.2 & 8.36  &8.97 &     11 &     12  \\
 54.0 --  75.0 &   62.6 & 1.61  &1.85 &      8.3 &     11  \\
 75.0 --  85.0 &   79.7 & 0.327 &0.384 &     11 &     10  \\
 85.0 -- 140.0 &  105.1 & 0.0414&0.0366 &    23 &     10  \\
\end{tabular}

\end{center}
\end{table}


\begin{references}

\bibitem{farrar}
G.R.~Farrar, Phys. Lett.{\bf B 67}, 337 (1977).

\bibitem{uatwo} 
UA2 Collaboration, J. Alitti {\it et al.,}
Phys. Lett. {\bf B 263}, 544 (1991).

\bibitem{cdf} 
CDF Collaboration,
F. Abe {\it et al.,} Phys. Rev. D~{\bf 48}, 2998 (1993);
Phys. Rev. Lett. {\bf 73}, 2662 (1994).

\bibitem{prl1a} D\O\ Collaboration, S. Abachi {\em et al.},
Phys. Rev. Lett.~{\bf 77}, 5011 (1996).

\bibitem{kt}
J. Huston {\it et al.,} Phys. Rev. D~{\bf 51}, 6139 (1995);\\
H.-L. Lai and H.-N. Li, Phys. Rev. D~{\bf 58}, 114020 (1998);\\
L. Apanasevich {\it et al.,} Phys. Rev. D~{\bf 59}, 074007 (1999).

\bibitem{vvv} 
M. Gl\"{u}ck, L.E.~Gordon, E.~Reya, and W.~Vogelsang,
Phys. Rev. Lett. {\bf 73}, 388 (1994);
W. Vogelsang and A. Vogt, Nucl. Phys. {\bf B 453}, 334 (1995).

\bibitem{dzero} D\O\ Collaboration, S. Abachi {\em et al.}, 
Nucl.\ Instrum.\ Methods {\bf A 338}, 185 (1994).

\bibitem{pythia} 
T. Sj\"ostrand, Comp. Phys. Commun. {\bf 82}, 74 (1994). 
Version 5.7 was used in this analysis.   

\bibitem{d0geant} 
J.~Womersley,~``The D\O\ Monte Carlo'', in Proc. of the 26th Int. Conf. on
High Energy Physics, Dallas, TX, 1992, Fermilab-Conf-92-306. 

\bibitem{hmcmll} 
R. Barlow and C. Beeston, Comp. Phys. Commun. {\bf 77}, 219 (1993). 

\bibitem{fragerror}
Suggested as an estimate of uncertainty in the neutral meson
multiplicity, based on comparisons with LEP data 
(T. Sj\"ostrand, private communication). 

\bibitem{bantly}
J. Bantly {\it et al.}, Fermilab-TM-1930, 1996 (unpublished).
In order to facilitate comparison with earlier data, 
this analysis does not use the luminosity normalization 
given in D\O\ Collaboration,  
B. Abbott {\it et al.}, hep-ex/990625, sec. VII, pp. 21-22 
(submitted to Phys. Rev. D). The updated normalization 
would have the effect of increasing the luminosity by 3.2\%.

\bibitem{owens} H. Baer, J. Ohnemus, and J.F. Owens,
Phys. Rev. D~{\bf 42}, 61 (1990).

\bibitem{CTEQ} CTEQ Collaboration, H.L.~Lai {\em et al.},
Phys. Rev. D~{\bf 55}, 1280 (1997).

\bibitem{e706} E706 Collaboration,
L. Apanasevich {\it et al.,} Phys. Rev. Lett. {\bf 81}, 2642 (1998). 

\end{references}
\end{document}